\documentclass[a4paper,fleqn]{cas-sc}

\usepackage[authoryear,longnamesfirst]{natbib}
\usepackage{lineno}

\def\tsc#1{\csdef{#1}{\textsc{\lowercase{#1}}\xspace}}
\tsc{WGM}
\tsc{QE}

\begin{document}
\let\WriteBookmarks\relax
\def\floatpagepagefraction{1}
\def\textpagefraction{.001}

\shorttitle{Lunar evolution with a magma ocean and fossil figure}    

\shortauthors{B.G. Downey et al.}  

\title [mode = title]{The thermal-orbital evolution of the Earth-Moon system with a subsurface magma ocean and fossil figure}  

\author[1]{Brynna G. Downey}
\cormark[1]
\ead{bgdowney@ucsc.edu}

\author[1]{Francis Nimmo}

\author[2]{Isamu Matsuyama}

\affiliation[1]{organization={Department of Earth and Planetary Sciences, University of California},
            city={Santa Cruz},
            postcode={95064}, 
            state={CA},
            country={USA}}
\affiliation[2]{organization={Lunar and Planetary Laboratory, University of Arizona},
            city={Tucson},
            postcode={85719}, 
            state={AZ},
            country={USA}}

\cortext[1]{Corresponding author}


\begin{abstract}
Various theories have been proposed to explain the Moon's current inclined orbit. We test the viability of these theories by reconstructing the thermal-orbital history of the Moon. We build on past thermal-orbital models and incorporate the evolution of the lunar figure including a fossil figure component. Obliquity tidal heating in the lunar magma ocean would have produced rapid inclination damping, making it difficult for an early inclination to survive to the present-day. An early inclination is preserved only if the solid-body of the early Moon were less dissipative than at present. If instabilities at the Laplace plane transition were the source of the inclination, then the Moon had to recede slowly, which is consistent with previous findings of a weakly dissipative early Earth. If collisionless encounters with planetesimals up to 140~Myr after Moon formation excited the inclination, then the Moon had to migrate quickly to pass through the Cassini state transition at 33~Earth radii and reach a period of limited inclination damping. The fossil figure was likely established before 16~Earth radii to match the present-day degree-2 gravity field observations.
\end{abstract}

\begin{highlights}
\item{We address the plausibility of origin theories for the Moon’s inclination}
\item{The Moon’s gravity field provides an additional constraint on the evolution} 
\item{Obliquity tidal heating in the magma ocean and crust damps out an early inclination}
\item{Two origin theories for the Moon’s inclination require different migration speeds}
\item{A fossil figure is established before 16 Earth radii}
\end{highlights}


\maketitle

\section{Introduction}\label{Sec:intro}
The origins of the Moon's present-day orbital inclination have long been a mystery \citep[e.g.,][]{macdonald1964tidal, goldreich1966history, mignard1981lunar, touma1994evolution}. In the canonical Moon-forming giant impact theory \citep[e.g.,][]{canup2001origin}, the Moon formed in the Earth's equatorial plane, i.e., a zero-inclination orbit, so the inclination had to arise during the Moon's outwards migration from Earth. In this work, we reconstruct the thermal-orbital history of the Moon to improve our understanding of how the Moon's orbit became inclined.

Various theories have been developed to explain excitations of the lunar inclination, which we categorize into early and late scenarios. Two early possibilities are that soon after the Moon-forming giant impact, resonances between the proto-Moon and the debris-disk excited the inclination to $15^\circ$ \citep{ward2000origin}, and passage through the evection and eviction resonances at 4.6-6~Earth radii excited the inclination up to 9-13$^\circ$ \citep{touma1998resonances}. A late possibility posits a high obliquity (60-80$^\circ$), high angular momentum early Earth that excited the inclination to $>30^\circ$ during the Moon's Laplace plane transition (LPT) at 16-22~Earth radii around 1-100~Myr after Moon formation \citep{cuk2016tidal,cuk2021tidal}. One last possibility is that during the period $10-100$~Myr after Moon formation, collisionless encounters with planetesimals gravitationally excited the inclination to its present value \citep{pahlevan2015collisionless}. 

Any viable inclination excitation has to contend with inclination damping due to tidal heating in the Earth and Moon \citep[e.g.,][]{chyba1989tidal}. In particular, \citet{chen2016tidal} found that obliquity tidal heating in the ancient lunar magma ocean \citep{tyler2008strong} would have accelerated the inclination decay while the obliquity grew to large values during the Cassini state transition (CST) at $\sim 30$~Earth radii. \citet{chen2016tidal} concluded that an early event can only be the source of the present-day inclination if the Moon's initial recession rate from the Earth were slow, the case for a weakly dissipative early Earth. This assumes that the magma ocean solidified in 100-200~Myr as supported by the lunar chronology data \citep[e.g.,][and references therin]{elkins2011lunar,maurice2020long}. If the early Earth was in fact dissipative, the inclination has to be from a late excitation mechanism. Our thermal-orbital model is based on \citet{chen2016tidal}, but to provide an additional constraint, we include the evolution of the lunar figure using the model in \citet{matsuyama2021lunar}. 

The lunar figure plays an important role in constraining the Moon's orbital evolution. Not only does it have a putative fossil figure, recording the orbital state at an earlier point in time \citep[e.g.,][]{jeffreys1915certain, lambeck1980lunar}, but it affects the obliquity through Cassini states. \citet{matsuyama2021lunar} developed a lunar figure model which includes a fossil figure component and incorporates the effects of obliquity and eccentricity on the shape. Their fossil figure estimates are consistent with establishing a fossil figure at $\sim 13$~Earth radii with obliquity $-0.16^\circ$ and eccentricity $<0.3$. This is in contrast with other works that tried to determine the state of the orbit when the fossil figure froze in, with most of them requiring either large eccentricities of $0.15 - 0.6$ \citep{garrick2006evidence, matsuyama2013fossil, keane2014evidence} or large semi-major axes of $>30$~Earth radii \citep{garrick2014tidal, qin2018formation}. The range of results stems from different methods for estimating the size of the fossil figure from the present degree-2 gravity field. For example, \citet{keane2014evidence} subtract contributions from mass anomalies, \citet{garrick2014tidal} subtract contributions from compensated topography, and \citet{matsuyama2021lunar} subtract the South Pole-Aitken basin contribution. Finally, \citet{qin2018formation} differ because they freeze the fossil figure in gradually over 500~Myr, but they use the fossil figure estimates in \citet{keane2014evidence}.

Other works besides \citet{chen2016tidal} have tried to constrain aspects of the Moon's thermal-orbital history, but they did not focus on the origins of the lunar inclination. First, \citet{meyer2010coupled} focused on whether the orbital parameters proposed in \citet{garrick2006evidence} to explain the fossil figure were plausible. Second, \citet{tian2017coupled} focused on whether the evection resonance could extract excess angular momentum from an early fast-spinning Earth. Third, \citet{daher2021long} developed a sophisticated framework for tides in Earth's oceans to evolve the lunar orbit, but they do not include the effects of the lunar magma ocean or a fossil figure on the evolution of the Earth-Moon system.

In this paper, we use our coupled thermal-orbital model to match the present-day inclination and degree-2 gravity observations to answer the questions of how the Moon got its inclination and where the Moon's fossil figure froze in. We build on the thermal-orbital model in \citet{chen2016tidal} and include a fossil figure component in the evolution of the lunar figure as in \citet{matsuyama2021lunar}. We also include (1) crustal tidal heating using a combination of a constant $k_2/Q$ and the model from \citet{garrick2010structure} (2) the effect of an overlying solid shell on magma ocean dissipation and (3) the eccentricity evolution. In Section~\ref{Sec:thermal-orbital evolution}, we describe the components of our thermal-orbital model. In Section~\ref{Sec:results} we find that a lasting early inclination requires that the Moon be more dissipative at present than in its past, and we present the necessary conditions to have a late inclination excitation as in \citet{pahlevan2015collisionless} or \citet{cuk2016tidal}. In Section~\ref{Sec:discussion} we discuss the importance of solid-body tides and the connection between the lunar fossil figure and the CST, and in Section~\ref{Sec:conclusion} we summarize our results and discuss plans for future work.

\section{Thermal-orbital evolution}
\label{Sec:thermal-orbital evolution}
We build a coupled thermal-orbital model based on \citet{chen2016tidal} and \citet{matsuyama2021lunar} that evolves the lunar orbit and lunar figure with time while tracking the solidification of the magma ocean. The goal is to find the conditions consistent with an early compared to late inclination excitation all while reproducing the observations of the present-day lunar figure.

\subsection{Orbital evolution}
The Moon's orbital evolution is governed by tidal dissipation in the Earth and Moon. Tidal dissipation in the Earth causes the Moon's semi-major axis and eccentricity to increase and the inclination to decrease whereas tidal dissipation in the Moon causes the semi-major axis, eccentricity, and inclination to all decrease. We use the Mignard equations to evolve the Earth-Moon system under the influence of tidal effects from the Earth, Moon, and Sun \citep{mignard1979evolution, mignard1980evolution, mignard1981lunar}. The Mignard equations track the Moon's semi-major axis, inclination, and eccentricity, and the Earth's obliquity and spin rate. See \citet{meyer2010coupled} and \citet{chen2016tidal} (includes corrected typos in \citet{meyer2010coupled}) for the Mignard equations and relevant terms. Their underlying principle is that there is a time lag, $\Delta t$, between when the perturbing body is directly overhead and when high tide on the body of interest occurs \citep{mignard1979evolution}. A larger $\Delta t$ means greater tidal dissipation and therefore greater rates of change of the orbital elements.

The $\Delta t$ of the early Earth is important for constraining the lunar evolution, but the past dissipative properties of the Earth are unknown and impossible to infer because dissipation in the Earth is currently dominated by ocean tides. Currently, $\Delta t \approx 600$~s \citep{munk1960rotation}, which if constant, implies the Moon formed $\sim$1~Ga. Therefore, $\Delta t$ had to be much smaller earlier on. Adding to our uncertainty is the fact that dissipation in the Earth also depends on the Earth's $k_2$ Love number at the tidal forcing frequency, $k_{2,E}$, which would have been larger in the past for a warmer, molten Earth. Because of this, we fold any uncertainty in $k_{2,E}$ into that for $\Delta t$ and treat $k_{2,E} \Delta t$ as a combined quantity. For $k_{2,E}=0.3$ at the present, $k_{2,E}\Delta t \approx 180$~s, while according to our model, the average $k_{2,E}\Delta t$ over the past $\sim$4.5~Gyr is $\sim$50~s. Dissipation in the early Earth could have been weak because at the higher frequencies associated with the Earth's faster rotation, the ocean normal modes are less well-matched with the degree-2 tidal forcing \citep{bills1999lunar}. Weak dissipation and thus a small $k_{2,E}\Delta t$ of the early Earth is further supported by \citet{zahnle2015tethered} who suggest that the thermal blanketing effect of the Earth's early atmosphere would have limited tidal heating. Two recent studies on the ocean dynamics of the Earth seem promising, but we do not incorporate their findings yet because both are incompatible with the Moon-forming giant impact theory; \citet{tyler2021tidal} and \citet{daher2021long} predict that the Moon formed at 44$R_E$ $\sim$4.5~Ga. \citet{daher2021long} acknowledge this with the intention of focusing on early sources of tidal dissipation in future works.

We create a step function for $k_{2,E}\Delta t$, which, given the uncertainties discussed above, is a simple parameterization that allows us to start the Earth with weaker dissipation then at a specific point in time switch to the enhanced tidal dissipation observed in the oceans today. To impose different orbital migration rates for the early Moon, we vary $(k_{2,E}\Delta t)_0$, and to keep the final semi-major axis accurate to within 10 per cent, we set $(k_{2,E}\Delta t)_f=180$~s at 1.25~Ga. Tidally-laminated sediment data and Earth ocean tide modelling are consistent with the high rate of Earth dissipation beginning $\sim$1~Ga \citep{bills1999lunar}. We have also experimented with smoothly-varying (exponential) descriptions of $k_{2,E}\Delta t$ and find very similar results to those presented below.

One critique of a constant $\Delta t$ formulation is that it is not consistent with the Moon's tidal response as observed in the GRAIL data \citep{williams2015tides}. However, despite the GRAIL data, it is still inconclusive which tidal model is most accurate and whether the lunar tidal response can be extrapolated to other planetary bodies \citep{williams2015tides}. Whatever the case, \citet{touma1994evolution} found that various tidal formulations, including the Mignard equations, yield approximately the same evolution for the Moon over billion year timescales because what matters most is accounting for all tidal effects. In light of this, we prioritize evolving the lunar orbit to the present-day with a step function $k_{2,E}\Delta t$ model for the Earth's tidal dissipation. For the Moon, we calculate $Q$ based on models for dissipation in the magma ocean and solid crust as described in Secs~\ref{Sec:solid-body tides} and \ref{Sec:ocean tides}.

\subsection{Cassini states and lunar figure}
\label{Sec:CSR}
We assume that the Moon is always in a Cassini state, namely that the spin pole precesses about the orbit normal at the same rate as the orbit normal precesses about the normal to the Laplace plane \citep[e.g.,][]{colombo1966cassini,peale1969generalized, ward1975past}. The Laplace plane is the average precessional plane, which for the Moon is currently the ecliptic, the Earth's orbit plane. Torques on the lunar figure cause the spin pole to precess, and torques from the Earth's oblateness and the Sun cause the orbit pole to precess. These two rates of precession are equal only when the spin axis is tilted by the Cassini state obliquity. We expect all satellites damped to synchronous rotation to also be in a damped Cassini state \citep{ward1975past}. In general there are four possible Cassini states. Close to the Earth, the Moon is in state 1, where the obliquity lies close to the orbit normal, and at around $34 R_E$, it transitions to state 2, where the obliquity lies close to the ecliptic \citep[e.g.,][]{ward1975past}. During the transition from state 1 to state 2, the magnitude of the obliquity increases rapidly which could lead to strong obliquity tidal heating and inclination damping. The lunar figure affects the spin pole precession and therefore the Cassini state obliquity, so it is important to have an accurate model of the size of the Moon's fossil and tidal-rotational bulges. The formulation in \citet{matsuyama2021lunar} includes the effects of obliquity and eccentricity on the Moon's deformation, quantified by the degree-2 gravity coefficients, $J_2$ and $C_{22}$ \citep{kaula1964tidal}, and allows for both a fossil and a tidal-rotational component.

Equating the precessional torques on the lunar orbit to those on the lunar figure gives the Cassini state relation provided in Eq.~\ref{Eq:csr} \citep{peale1969generalized, bills2008forced, bills2011rotational}. A satellite's obliquity, $\theta_0$, can be numerically solved for given various gravitational and orbital parameters. The Cassini state relation and the orbit precession rate \citep[e.g.,][]{goldreich1966history} are as follows:
\begin{flalign}
    \label{Eq:csr}
    \frac{3}{2} & \left[\left(J_2+C_{22}\right)\cos \theta_0+C_{22}\right]p\sin \theta_0 = c\sin \left( i-\theta_0 \right) \\
    \label{Eq:omegadot}
    \dot{\Omega}_{\text{orb}} &= -\frac{3}{2} n J_{2,E}\left(\frac{R_E}{a}\right)^2 - \frac{3}{4} n \frac{M_S}{M} \left(\frac{a}{a_E}\right)^3 \\
    J_{2,E} &= \frac{1}{3}k_{2,E}\left(\frac{\omega}{n_G}\right)^2. 
\end{flalign}
Here, $a$, $n$, $i$, and $c$ are the lunar semi-major axis, mean motion, inclination, and normalized polar moment of inertia, $p = n/\dot{\Omega}_{\text{orb}}$, where $\dot{\Omega}_{\text{orb}}$ is the rate of nodal regression, $M$, $R_E$, $J_{2,E}$, $k_{2,E}$, $a_E$, and $\omega$ are the Earth's mass, radius, $J_2$, long-term Love number, heliocentric semi-major axis, and spin rate, $\displaystyle n_G = (GM/R_E^3)^{1/2}$ is the grazing mean motion about the Earth, and $M_S$ is the Sun's mass. The first term in Eq.~\ref{Eq:omegadot} is the torque from the Earth's oblateness on the lunar orbit, and the second term is the solar torque in the limit that $a \ll a_E $. The LPT happens when the solar torque becomes stronger than the Earth's, so when the second term in Eq.~\ref{Eq:omegadot} is larger in magnitude than the first. The CST happens when the Moon's spin pole becomes closer to the Laplace plane pole than to the orbit normal, so when the numerically-solved for obliquity from Eq.~\ref{Eq:csr} first flips from being negative (state 1) to positive (state 2). 

The degree-2 gravity coefficients associated with the rotational and tidal potentials averaged over the orbital and precession periods are given by \citep{matsuyama2021lunar}
\begin{flalign}
    <J_2> & = q^T \left[ \frac{1}{3}+\frac{1}{32}\left(2-3\sin^2\theta_0\right)\left(8+12e^2+15e^4\right)\right] \\
    <C_{22}> & = q^T \frac{1}{256}\left(16 - 40e^2 + 13e^4 \right) \left(1+\cos \theta_0 \right)^2 \\
    q^T & = \left(\frac{M}{m}\right)\left(\frac{R}{a}\right)^3,
\end{flalign}
where $m$, $R$, and $e$ are the Moon's mass, radius, and orbital eccentricity. Using these equations and assuming that a fossil ﬁgure is established as an elastic lithosphere forms over a timescale much shorter than the orbital evolution timescale, the degree-2 gravity coefficients of the Moon including the fossil figure contribution can be written as 
\begin{flalign}
    \label{Eq:J2_fossil}
    J_2 & = (k_2^{\infty*}-k_2^\infty)<J_2^*> + k_2^\infty<J_2> \\
    \label{Eq:C22_fossil}
    C_{22} & = (k_2^{\infty*}-k_2^\infty)<C_{22}^*> + k_2^\infty<C_{22}>;
\end{flalign}
asterisks denote the values when the fossil figure freezes in, and $k_2^\infty$ is the degree-2 long-term tidal Love number, which is a function of the lunar elastic lithosphere thickness. 
In this work we approximate the elastic lithosphere thickness by the crustal thickness, whose growth due to magma ocean solidification we describe in more detail below in Section~\ref{Sec:magma ocean solidification}. For greater elastic thicknesses, $k_2^\infty$ will be smaller indicating a more rigid body. Instantaneously establishing a fossil figure is simplistic compared to the long-term visco-elastic process in \citet{qin2018formation}, but out results differ by $\lesssim 25$ per cent.

We compute the long-term tidal Love number $k_2^\infty$ using the classical propagator matrix method (e.g., \citet{Sabadini2016love}) and assuming a 4-layer interior structure consisting of a liquid core, mantle, magma ocean, and shell with the interior structure parameters summarized in Table~\ref{tab:Moon params}. As the magma ocean solidifies as described below, the core density is computed self-consistently so as to satisfy the mean density constraint assuming a 380~km core radius. We assume a fossil ﬁgure preserved by an elastic lithosphere because this is the ﬁrst region to develop long-term elastic strength as the Moon cools. Therefore, the rigidity of all layers is set to zero except for the elastic lithosphere.

The term $(k_2^{\infty*}-k_2^\infty)$ in equations~\ref{Eq:J2_fossil} and \ref{Eq:C22_fossil} measures the relative rigidity of the Moon when the fossil figure freezes in $(k_2^{\infty*})$ to that throughout the simulation $(k_2^\infty)$ in order to determine how well the lithosphere supports the stresses of the fossil figure at any point in time. The more rigid the Moon becomes from a growing elastic thickness (smaller $k_2^\infty$), the less it will deform in the new rotational and tidal potential, and the more it will revert back to the fossil figure (larger $k_2^{\infty*}-k_2^\infty$). The fossil figure is the shape that without any external forces the body would adopt. We do not account for any long-term relaxation of the fossil stresses.

Where the fossil figure froze in is a variable in our model. Comparing the model results to the degree-2 gravity observations in \citet{matsuyama2021lunar} allows us to provide an estimate of the orbital state when the fossil figure froze in.

\subsection{Tidal heating}
Tidal heating in the Moon damps the inclination, eccentricity, and semi-major axis and is the key obstacle to the survival of an early inclination. The mechanisms for tidal dissipation in the solid-body and magma ocean are different, so we provide details on each below. 

\subsubsection{Solid-body tidal heating}
\label{Sec:solid-body tides}
For solid-body tidal heating in the lunar crust, we use a combination of a visco-elastic model that depends on the Moon's Love numbers, viscosity, and rigidity and a constant $k_2/Q$ model, where $k_2$ is the Moon's degree-2 potential Love number at the tidal forcing frequency and $Q$ is the Moon's tidal quality factor. The visco-elastic model is used while the magma ocean is still present since $Q$ could vary by several orders of magnitude as the crust and mantle are cooling. After magma ocean solidification, using just the visco-elastic model and a crustal viscosity of $\eta=10^{19}$~Pa~s produces values for $Q$ on the order of $10^6$-$10^8$, which would underestimate solid-body tidal heating at the present. Because of this, once the magma ocean is solidified, we switch to calculating solid-body tidal dissipation from a constant $k_2/Q=6\times10^{-4}$, which is the value at the present-day \citep{williams2015tides}. A constant $k_2/Q$ balances the competing effects of the Moon behaving more elastically at the higher tidal frequencies associated with smaller semi-major axes with the Moon becoming less dissipative as it cools after the solidification of the magma ocean.

For the visco-elastic model, we calculate the heating rate from \citet{garrick2010structure} and apply it in a thin layer above the magma ocean to crudely approximate tidal heating in a temperature-dependent medium. Note that we neglect dissipation in the deep interior beneath the magma ocean, which if present, would further increase eccentricity and inclination damping. Eccentricity and obliquity tides create a tidal strain rate in the lunar crust, $\dot{\epsilon}$, which results in a volumetric tidal dissipation rate, $W$, whose equations are reproduced below from \citep{garrick2010structure}:
\begin{flalign}
    \dot{\epsilon}_e & = f\left(\frac{h_2}{2.5}\right)\frac{n^3R^3}{Gm}e \\
    W_e & = \frac{2\dot{\epsilon}_e^2 \eta}{1+\left(n \eta / \mu\right)^2} \\
    \label{eq:W}
    W & = W_e \left( 1 + \frac{\sin^2{\theta_0}}{7 e^2} \right).
\end{flalign}
The constant $f$ quantifies spatial variations in the strain rate and is of order unity, $h_2$ is the degree-2 displacement Love number at the tidal frequency, $G$ is the gravitational constant, and $\eta$ and $\mu$ are the viscosity and rigidity of the lunar crust. The eccentricity tidal strain rate, $\dot{\epsilon}_e$, has a corresponding heating rate, $W_e$ \citep{ojakangas1989thermal, garrick2010structure}. To account for obliquity tides in the full heating rate, $W$, we approximate the ratio of eccentricity tidal dissipation to that of obliquity tides to be $\displaystyle \sin^2{\theta_0}/7e^2$. This is based on the standard tidal dissipation expressions in \citet{peale1978contribution} and \citet{peale1980tidal}.

We compute $h_2$ and $k_2$ at the tidal forcing frequency using the propagator matrix method and the correspondence principle \citep[e.g.,][]{Tobie2005lovenumbers}. For this computation, we assume the 4-layer interior structure described above, a Maxwell rheology, a mantle with a rigidity of 70 GPa and a viscosity of $10^{21}$ Pa s, a magma ocean, and a shell with a rigidity of 30~GPa and a viscosity of $10^{21}$ Pa s. The visco-elastic Love numbers are not sensitive to the assumed viscosities because the Maxwell time of the solid layers is significantly larger than the tidal forcing period, corresponding to a nearly purely elastic response that is independent of the assumed viscosity and the forcing frequency.

Most of the crust is likely to be too cold and rigid to dissipate much energy, so we concentrate the dissipation in a layer of thickness $\delta$ at the bottom of the crust of total thickness $d$ overlying the magma ocean. The thickness of this dissipative layer depends on the temperature profile and heat production of the crust and the conductive cooling of the magma ocean. Because all the layers considered are thin compared to the Moon's radius, a Cartesian geometry is appropriate.

For an unheated region, $(z < d-\delta)$, where $z$ is depth into the crust, the temperature structure increases linearly with depth due to conductive cooling from the interior to the surface:
\begin{equation}
    T(z) = T_0 + \frac{T_a-T_0}{d-\delta}z,
\end{equation}
where $T_0$, $T_m$, and $T_a$ are the temperatures of the surface, magma ocean (i.e., the base of the crust), and top of the dissipative layer. The interface temperature $T_a$ at $z=d-\delta$ is determined by the sensitivity of the viscosity to temperature via the activation energy. For silicates, this $e$-folding temperature drop is roughly 50~K.

For an internally heated region, $(z > d-\delta)$, the solution to the temperature structure has a similar linear component to conduct heat from the magma ocean to the surface but includes terms to account for heat generation and diffusion in this layer:
\begin{equation}
    T(z) = T_m
            -\frac{1}{\delta}\left(T_m - T_a\right)\left(d-z\right)
            +\frac{W}{2k}
            \left[\delta\left(d-z\right) - (d-z)^2\right],
\end{equation}
where $k$ is the thermal conductivity and $W$ is the internal volumetric heating rate from Eq.~\ref{eq:W}. For simplicity, we assume that $W$ is constant within the heated layer because the viscosity variation within this layer is small. It can be verified that this temperature profile gives the correct limits of $T_m$ at $z=d$ and $T_a$ at $z=d-\delta$. 

If we specify the basal heat flux $H_b$, then the heat flux balance at the surface is
\begin{equation}
    \label{Eq:heatflux2}
    W\delta+H_b = k\frac{T_a-T_0}{d-\delta},
\end{equation}
and the heat flux balance in the heated region is
\begin{equation}
    \label{Eq:heatflux1}
    \frac{W\delta}{2}+H_b = k \frac{T_m-T_a}{\delta}.
\end{equation}

Since we know the crustal thickness, $d$, from solidification of the magma ocean and the internal volumetric heating rate, $W$, from tides, we can solve for the basal heat flux $H_b$ and $\delta$ simultaneously. Combining equations~\ref{Eq:heatflux2} and \ref{Eq:heatflux1} gives rise to a cubic relation that can be solved numerically:
\begin{equation}
    \label{delta}
    \frac{1}{2}W \delta^3-\frac{1}{2} W d \delta^2 + k\delta \left(T_m-T_0\right)-k d\left( T_m - T_a \right) = 0.
\end{equation}
In the limit that the heat flux from cooling of the magma ocean is much greater than the internally generated tidal heating $(H_b \gg W\delta)$, the thickness of the heated layer is proportional to the temperature contrast, $\displaystyle \delta \approx d (T_m-T_a)/(T_m-T_0)$ as expected.

The heat production rate in the crust is then $H_{\text{crust}} = W \delta$, and the total solid-body energy dissipation in the shell is $\dot{E}_{\text{sol}} = 4\pi R^2 H_{\text{crust}}$. This can be broken down into an eccentricity tide component, $\dot{E}_{e,\text{sol}}$ and obliquity tide component, $\dot{E}_{\theta,\text{sol}}$ in the following way:

\begin{flalign}
    \dot{E}_{e,\text{sol}} & =  \frac{7e^2}{7e^2+\sin^2\theta_0} \dot{E}_{\text{sol}} \\
    \dot{E}_{\theta,\text{sol}} & = 
    \frac{\sin^2\theta_0}{7e^2+\sin^2\theta_0} \dot{E}_{\text{sol}} \\
    \label{Eq:Qeff}    
    Q_{\text{eff}} & = \frac{3}{2}k_2\frac{n^5R^5}{G}\frac{\left(7e^2+\sin^2\theta_0\right)}{\dot{E}_{\text{sol}}},
\end{flalign}
where $Q_{\text{eff}}$ is the effective tidal quality factor to compare these tidal heating rates to the standard dissipation expressions in \citet{peale1980tidal}. Once the magma ocean is solidified, we use Eq.~\ref{Eq:Qeff} and a constant $k_2/Q$ to calculate solid-body tidal heating. This equation assumes small eccentricity and obliquity \citep[c.f.,][]{wisdom2008tidal} and so during the CST underestimates solid-body obliquity heating by a factor of 1.5-3 at the peak obliquity, which does not qualitatively change our results.

\subsubsection{Magma ocean tidal heating}
\label{Sec:ocean tides}
Tidal heating in the lunar magma ocean was a novel addition to the thermal-orbital model in \cite{chen2016tidal} and was found to play one of the most important roles in the lunar evolution. A resonance in the ocean flow of synchronous satellites between the obliquity tide and a Rossby-Hauritz wave leads to enhanced dissipation \citep{tyler2008strong}. \citet{chen2014tidal} showed that obliquity tidal heating in oceans on synchronous satellites can often be stronger than that in the solid-body, so neglecting ocean tides risks drastically underestimating inclination damping. 

We use the analytical expressions in \citet{hay2019subsurface} to calculate obliquity and eccentricity tidal heating in the lunar magma ocean due to bottom drag. These expressions include the effect of a rigid lid overlying the ocean, which is important as the magma ocean crystallizes and the crust thickens. The effect of pressure forcing on the ocean from the lid is included using pressure Love numbers \citep{matsuyama2018subsurface}. We compute the pressure Love numbers using the propagator matrix method with the appropriate boundary conditions for pressure forcing \citep{matsuyama2018subsurface} and under the same assumptions described above for the tidal Love numbers at the tidal forcing frequency. In our baseline models, the bottom drag coefficient has the same value as for Earth's oceans, $c_D = 0.002$.

\subsection{Magma ocean solidification}
\label{Sec:magma ocean solidification}
After the Moon-forming giant impact, the Moon would have had a global magma ocean 100–1000~km deep. We track the solidification of the magma ocean and growth of an overlying crust due to conductive cooling and allow for remelting due to tidal dissipation. The magma ocean lifetime depends chiefly on the thermal diffusivity, $\kappa = k/\rho_o C_p$, where $C_p$ is the specific heat capacity of the magma ocean. 

The lifetime of the lunar magma ocean is estimated from the span of ages inferred for crustal formation, 100-200~Myr \citep[e.g.,][and references therein]{elkins2011lunar, maurice2020long}, although there are surface ages with high uncertainties \citep{borg2015review}. \citet{maurice2020long} found that the last dregs of the lunar magma ocean solidified 100-180~Myr after the most reliably dated sample from the earliest flotation crust \citep{borg2011chronological}. A magma ocean lifetime of 100-200~Myr can be attained by using the thermal conductivity of anorthosite, $k=1.5$~W m$^{-1}$ K$^{-1}$ \citep{maurice2020long}.

Of a layer that solidifies in the magma ocean, approximately 80 per cent solidifies at the bottom of the ocean and 20 per cent floats to the crust \citep{snyder1992chemical,warren1986bulk}. Therefore, 20 per cent of the change in the ocean thickness goes to or comes from the Moon's crust. The rate of change for the magma ocean and crustal thicknesses, $h$ and $d$, are as follows:
\begin{flalign}
    \rho_o L \frac{dh}{dt} & = - \frac{k\left(T_a-T_0\right)}{d-\delta} + H_{\text{crust}} + H_{\text{ocean}} \\
    \dot{d} & = -\frac{1}{5}\frac{dh}{dt},
\end{flalign}
where $\rho_o$ is the magma ocean density, and $L$ is the latent heat of fusion. The ocean heat flux is $H_{\text{ocean}} = \dot{E}_{\text{ocean}}/4\pi R^2$, where $\dot{E}_{\text{ocean}}$ is tidal dissipation in the subsurface lunar magma ocean calculated from \citet{hay2019subsurface}.

The depth of the magma ocean affects eccentricity tidal heating, but for small $c_D$, obliquity tidal heating is independent of ocean thickness \citep{hay2019subsurface}. The thickness of the crust affects solid-body tidal dissipation, and since we approximate the elastic lithosphere thickness by the crustal thickness, it also affects how well the Moon can maintain the lithospheric stresses of a fossil figure.

\subsection{Summary}
\begin{figure}
    \centering
{\includegraphics[width=0.8\columnwidth]{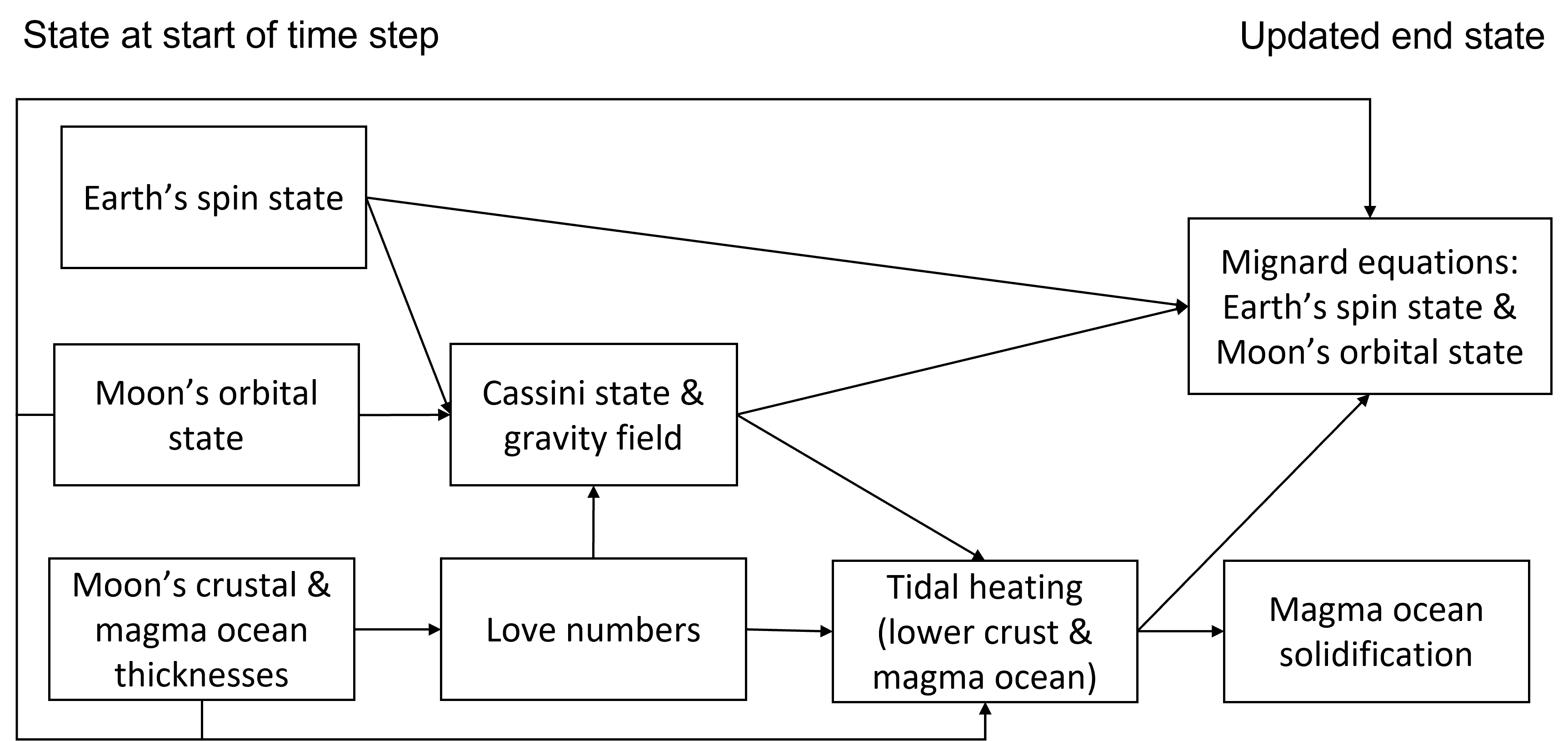}}%
        \caption{Block diagram of how the thermal and orbital processes relevant to the lunar inclination interact and depend on each other in each time step. The Earth's spin state, the Moon's orbital state, and the properties of the lunar crust all affect the Moon's Cassini state obliquity. Obliquity tidal heating in the solid-body and magma ocean then damp the inclination.}%
    \label{fig:block diagram}%
\end{figure}

In summary, the thermal-orbital model evolves the lunar orbit in response to tidal dissipation in the Earth and in the lunar magma ocean and lower crust. The biggest unknowns are the early migration rate of the Moon away from the Earth and where the Moon's fossil figure freezes in. These quantities affect whether the magma ocean solidifies before the LPT and whether the CST occurs before the planetesimal population dies out. A block diagram of how all of the relevant processes affect each other is included in Figure~\ref{fig:block diagram}.

We start the Moon at $a=6.5 R_E$, outside the evection and eviction resonances with a variable initial inclination to the Laplace plane, $i_0$. To test an early inclination, $i_0 = 12^\circ$, a plausible value after proto-Moon disk resonances \citep{ward2000origin} or the eviction resonance \citep{touma1998resonances}. To test a late inclination, $i_0=0^\circ$. To get the Earth approximately to its present-day state, the Earth's initial day is 5.6~hr, and its obliquity with respect to the Laplace plane is $I_0 = 8^\circ$. For a 1000~km magma ocean after Moon formation, a flotation crust does not form until the ocean solidifies to 200~km, so we set the initial ocean thickness as $h_0 = 200$~km and the initial crustal thickness as $d_0 = 1$~km. Tidal heating in the magma ocean is shut off at $h=10$~km to account for non-uniform solidification of the global magma ocean.

\section{Results}
\label{Sec:results}
We tested three inclination excitation scenarios in the context of our thermal-orbital model and determined which configurations matched the present-day observations for the lunar inclination and degree-2 gravity field.

The first scenario that we tested is that the Moon acquired an inclination of $12^\circ$ soon after Moon formation either through resonances between the proto-Moon and accretion disk \citep{ward2000origin} or from passage through the evection and eviction resonances \citep{touma1998resonances}.

The second scenario from \citet{cuk2016tidal} is that post-Moon-forming giant impact, the Earth-Moon system had a much higher angular momentum and obliquity ($60-80^\circ$) than it has now. The high angular momentum would allow for mixing of the proto-Earth and impactor material and explain the isotopic similarities between the Earth and Moon. \citet{tremaine2009satellite} found that for planets with high obliquity, a satellite's LPT, where the Sun's gravitational effect on the precession of the lunar orbit outweighs that from the planet's oblateness, can destabilize its orbit and add significant inclination and eccentricity. In \citet{cuk2016tidal, cuk2021tidal}, solar perturbations and resonances during the Moon's LPT drain excess angular momentum, diminish the Earth's large obliquity, and excite the lunar inclination to $>30^\circ$.

The third scenario taken from \citet{pahlevan2015collisionless} is that planetesimals from the inner solar system would have swept by the Earth-Moon system. A few bodies totalling 0.015~Earth mass would have collided with the Earth, being part of its late accretion. As the planetesimals swept through, collisionless encounters with the Moon imparted momentum and stochastically increased the Moon's inclination and eccentricity until the population died out around 140~Myr after Moon formation.

\begin{figure}
    \centering
{\includegraphics[width=\columnwidth]{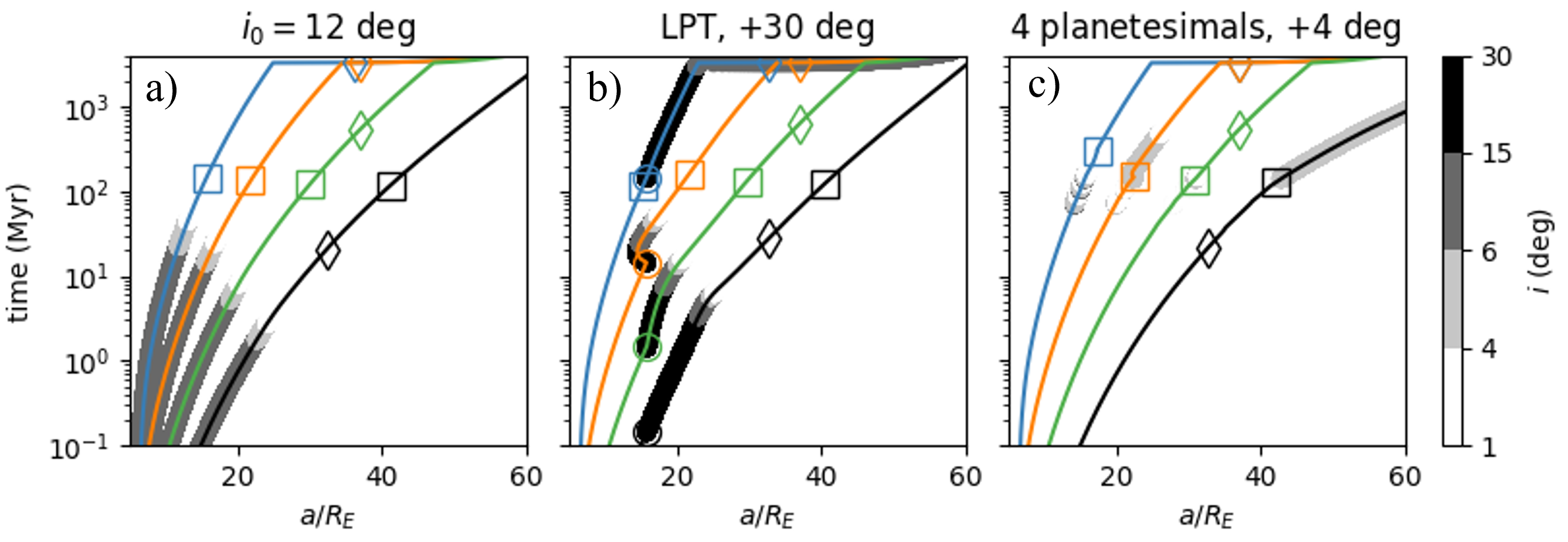}}%
        \caption{Three example inclination excitation scenarios. a) an early excitation with $i_0=12^\circ$ b) a $30^\circ$ excitation at the LPT c) four planetesimals adding $4^\circ$ of inclination at 35, 70, 105, and 140~Myr. The inclination evolution is shown in the greyscale contours, and the different lines are for four $(k_{2,E}\Delta t)_0$, 0.1~s (blue), 1~s (orange), 10~s (green), 100~s (black). Squares mark the lunar magma ocean solidification, diamonds the CST, and circles the LPT. The thermal diffusivity of the magma ocean is $\kappa=5\times 10^{-7} m^2 s^{-1}$, and it solidifies in 100-200~Myr. Once the magma ocean is solidified, the solid-body tidal dissipation is calculated using $k_2/Q=6\times 10^{-4}$, the present value. The knee in the graphs at 3.25~Gyr is when $k_{2,E}\Delta t$ steps to the present value of 180~s. Under no condition does an early inclination survive. In panel (b), for $(k_{2,E}\Delta t)_0 \le 0.1$~s, the magma ocean solidifies prior to the LPT, limiting ocean obliquity tidal heating and leaving a lasting inclination. In panel (c), for $(k_{2,E}\Delta t)_0 \ge 40$~s, the CST occurs and the magma ocean solidifies prior to 140~Myr when the planetesimal population dies down, allowing planetesimal perturbations to leave a lasting inclination.}%
    \label{fig:inc-scenarios}%
\end{figure}

Fig.~\ref{fig:inc-scenarios} shows example semi-major axis and inclination evolutions for a general early excitation ($<1$~Myr) and the two late excitation scenarios ($>1$~Myr). We varied the early migration rate of the Moon which is set by $(k_{2,E}\Delta t)_0$. To match the present-day degree-2 gravity field, each simulation has a different semi-major axis where the fossil figure freezes in, $a^*$. Fig.~\ref{fig:inc-scenarios}a shows the evolution for an early inclination excitation with an initial inclination to the Laplace plane of $i_0=12^\circ$. In no cases does the inclination survive to the present day. Fig.~\ref{fig:inc-scenarios}b  approximates a $30^\circ$ inclination excitation at the LPT, which for the angular momentum of the Earth-Moon system in our simulations happens at $a\sim16 R_E$. In this case the inclination survives if the outwards motion is sufficiently slow ($(k_{2,E}\Delta t)_0 \le 0.1$~s) that the LPT occurs after the magma ocean has solidified. Fig.~\ref{fig:inc-scenarios}c approximates perturbations to the lunar orbit from close encounters with planetesimals by having a set of four inclination kicks of $4^\circ$ at 35, 70, 105, and 140~Myr. In this case an inclination can survive if outwards motion is sufficiently rapid that the CST occurs prior to the disappearance of the planetesimals.  While Figs.~\ref{fig:inc-scenarios}b, c represent severe simplifications compared to the original works, they allow us to focus on the magnitude of the inclination excitation and whether it could survive tidal dissipation. As discussed more in-depth below, obliquity tidal dissipation is so strong that any early inclination damps, and the only way for a late inclination to survive is if the excitation occurs after strong ocean obliquity tidal heating ends.

\subsection{Early inclination excitation}
\label{Sec:early inc}
Fig.~\ref{fig:inc-scenarios}a shows that an early inclination cannot survive inclination damping due to obliquity tidal heating in the magma ocean and solid-body. Obliquity tidal dissipation in the magma ocean could be weaker for several reasons. The first is that the magma ocean could have been short-lived, although there is little evidence of this and would be contradictory to the lunar chronology data \citep[e.g.,][]{maurice2020long}. The second is that thickness variations in the overlying flotation crust could disrupt the resonance between the obliquity tide and the Rossby-Haurwitz wave \citep{rovira2020tides}. The final reason is that the strength of the tidal heating depends on the bottom drag coefficient, which for Earth is $c_D = 0.002$ and follows a scaling law relationship with the Reynolds number \citep{fan2019impacts}. The Reynolds number of the lunar magma ocean is calculated using the flow speed from \citet{chen2014tidal} to be self-consistent with our tidal heating model and the viscosity of basalt, $\eta = 1000$~ Pa s. The scaling law in \citet{fan2019impacts} produces values of $c_D$ in the lunar magma ocean in the range of 0.001-0.004, so we use $c_D=0.002$ for our baseline model.

\begin{figure}
    \centering
{\includegraphics[width=\columnwidth]{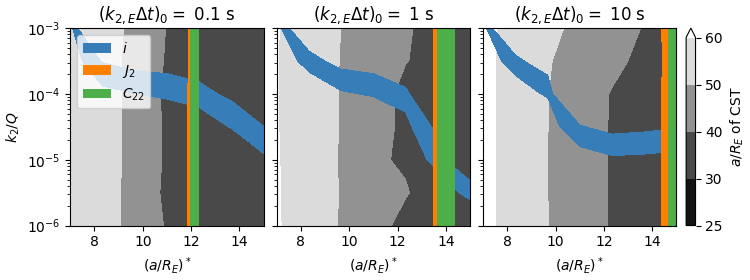}}%
        \caption{Parameters for which an early inclination survives. Including only solid-body and no ocean obliquity tides, for three different initial migration rates set by $(k_{2,E}\Delta t)_0$, the colored bands are the set of $k_2/Q$ and $a^*$ that match the present-day inclination (blue), $J_2$ (orange), and $C_{22}$ (green) to within 10 per cent. The $J_2$ and $C_{22}$ observations come from \citet{matsuyama2021lunar} and include the tidal-rotational and fossil figure components (excludes contributions from the South Pole-Aitken basin). The gray contours in the background show where the CST occurred in the model runs. To match the present-day $J_2$ and $C_{22}$ observations, the CST happened between 32 and 38 Earth radii. On average in the past, $k_2/Q$ would have to be at least one to two orders of magnitude lower than at present ($k_2/Q=6\times 10^{-4}$) for an early inclination to survive to the present-day.}%
    \label{fig:Moon-k2Q-contour}%
\end{figure}

Even if there were no ocean obliquity tidal heating, solid-body dissipation would still have to be weaker on average than at present to not damp an early inclination. Fig.~\ref{fig:Moon-k2Q-contour} shows the $a^*$ and average $k_2/Q$ over the past needed to allow an early inclination of 12$^\circ$ to survive and the present-day degree-2 gravity field to be reproduced. In this case, there are no ocean obliquity tides and the solid-body tides follow the visco-elastic model in Sec.\ref{Sec:solid-body tides} until solidification is complete in 100-200~Myr. The average $k_2/Q$ in the past would have to be $10^{-5}-10^{-4}$ for an early inclination of 12$^\circ$ to survive to the present-day while simultaneously satisfying the gravity constraints, which implies that the early Moon would have to be less dissipative than it is now. This is an upper bound on the needed average $k_2/Q$ because if there were any ocean obliquity tidal heating, the solid-body component would have to be weaker. A warmer, less rigid Moon is expected to be more dissipative, so we conclude that there are no conditions under which tidal dissipation can be weak enough to allow an early inclination of 12$^\circ$ to survive.

\subsection{Late inclination excitation}
\label{Sec:late inc}

\begin{figure}
    \centering
{\includegraphics[width=0.55
\columnwidth]{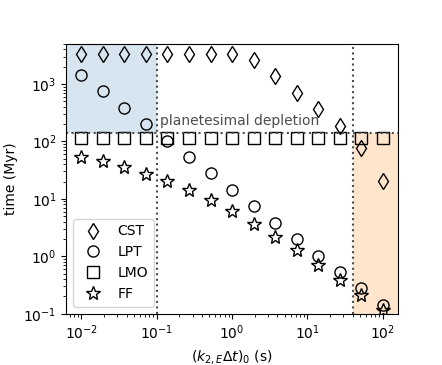}}%
        \caption{The timings of the fossil figure freeze-in (FF, star), the lunar magma ocean solidification (LMO, square), the LPT (circle), and the CST (diamond) as a function of $(k_{2,E}\Delta t)_0$. Each fossil figure semi-major axis is chosen for that run to match the $J_2$ and $C_{22}$ observations. If the LPT at $a\sim16 R_E$ is responsible for exciting the lunar inclination, then $(k_{2,E}\Delta t)_0 \le 0.1$~s to reach this point after the magma ocean solidifies in 100-200~Myr (represented by the blue shaded region in the upper left). If the planetesimal population is responsible, then $(k_{2,E}\Delta t)_0 \ge 40$~s to pass the CST and to solidify the magma ocean prior to the depletion of the planetesimals by 140~Myr (represented by the orange shaded region in the lower right).}%
    \label{fig:CST-LPT}%
\end{figure}

Obliquity tidal heating in the magma ocean and solid-body will damp an early inclination, so the only explanation for the present-day inclination is a late excitation mechanism that takes place after the period of strong ocean heating is over. Fig.~\ref{fig:CST-LPT} shows the relative timings of the LPT, CST, and the solidification of the lunar magma ocean as a function of $(k_{2,E}\Delta t)_0$ in order to determine what migration rates are needed for each of the late excitation mechanisms to be viable. For each $(k_{2,E}\Delta t)_0$, the fossil figure is established at a different semi-major axis so that the present-day degree-2 gravity field is reproduced. The blue shaded region in the upper left shows when the LPT postdates solidification of the magma ocean. The orange shaded region in the lower right shows when both the CST and the magma ocean solidification precede depletion of the planetesimal population at 140~Myr. From Fig.~\ref{fig:CST-LPT}, there is a maximum speed limit for instabilities at the LPT and a minimum speed limit for perturbations from plantesimals to be the source of the lunar inclination.

Instabilities at the LPT can only leave a lasting inclination if it occurs after the magma ocean solidifies otherwise ocean obliquity tidal heating damps the excitation in a span of 10's~Myr (see Fig.~\ref{fig:inc-scenarios}b). The LPT occurs when solar effects on the precession of the Moon's longitude of the ascending node outweigh effects from the Earth's oblateness. Quantitatively, this is when the second term in Eq.~\ref{Eq:omegadot} surpasses the first in magnitude. In our simulations, the LPT occurs consistently at $a\sim 16 R_E$ because we do not vary the initial angular momentum of the Earth-Moon system set by the initial lunar semi-major axis and spin rate of the Earth. The Earth's spin frequency evolution affects the Earth's oblateness quantified by the degree-2 gravity coefficient $J_{2,E}$. For the Moon to take at least 100~Myr to get to the LPT at $a\sim16 R_E$, there is a maximum speed limit, which translates to $(k_{2,E}\Delta t)_0 \le 0.1$~s (see Fig.~\ref{fig:CST-LPT}). Then to match the present-day degree-2 gravity observations in \citet{matsuyama2021lunar}, a fossil figure freezes in at $a^* \le 12 R_E$ (see Fig.~\ref{fig:Moon-k2Q-contour}). In reality, for the LPT to be destabilizing, the initial angular momentum of the Earth-Moon system would have to be higher than in our simulations \citep{cuk2016tidal}. This would push the LPT out to 17-22 $R_E$ \citep{cuk2021tidal}, raising the upper bound on the Moon's initial migration rate.

Collisionless encounters with planetesimals can only leave a lasting mark on the inclination if the magma ocean solidifies and the CST occurs before the depletion of the planetesimal population at $\sim 140$~Myr after Moon formation \citep{pahlevan2015collisionless}. The CST marks when the obliquity flips from negative to positive in the simulation (see Section~\ref{Sec:CSR}). Fig.~\ref{fig:inc-scenarios}c shows that because the planetesimal kicks are smaller than that from the LPT, even solid-body obliquity tides during the CST could damp out any planetesimal inclination excitation. Counterintuitively, planetesimal kicks too large early on make it harder for later kicks to survive because the tidal heating from the increased inclination and obliquity delays the magma ocean solidification and prolongs strong inclination damping. There is a minimum speed limit for the Moon to pass the CST before 140~Myr, which places the constraint $(k_{2,E}\Delta t)_0 \ge 40$~s (see Fig.~\ref{fig:CST-LPT}). A fossil figure freezes in at $a^* \sim 15 R_E$ to match the present-day gravity field observations (see Fig.~\ref{fig:Moon-k2Q-contour}).

Predicting which scenario is responsible for the lunar inclination depends largely on the $(k_{2,E}\Delta t)_0$ of the early Earth. \citet{zahnle2015tethered} found that after the Moon-forming giant impact, the Earth's atmosphere cooled slowly enough to limit dissipation in the interior. They predict $Q \gtrsim 10^4$, which by the tidal relation $1/Q \sim 2 \Omega \Delta t$, corresponds to $(\Delta t)_0 \lesssim 0.2$~s assuming the 5.6~hr initial Earth length of day used in this work. For a warmer young Earth, $k_{2,E}$ would be of order unity yielding $(k_{2,E}\Delta t)_0 \lesssim 0.2$~s. This constraint is consistent with the LPT excitation mechanism. Collisionless encounters with planetesimals could only impart inclination to a slowly receding Moon if it were less dissipative in the past than at present (see Sec.~\ref{Sec:early inc} for a discussion on the ways to decrease tidal dissipation in the Moon). A further difficulty is that collisionless encounters have a stronger effect at greater Earth-Moon distances, so slow lunar recession keeps the Moon at shorter distances while the planetesimal population depletes, exciting the inclination less \citep{pahlevan2015collisionless}. On the whole, therefore, we favour an early slow outwards migration and inclination excitation via instabilities at the LPT.

\begin{figure}
    \centering
{\includegraphics[width=0.9\columnwidth]{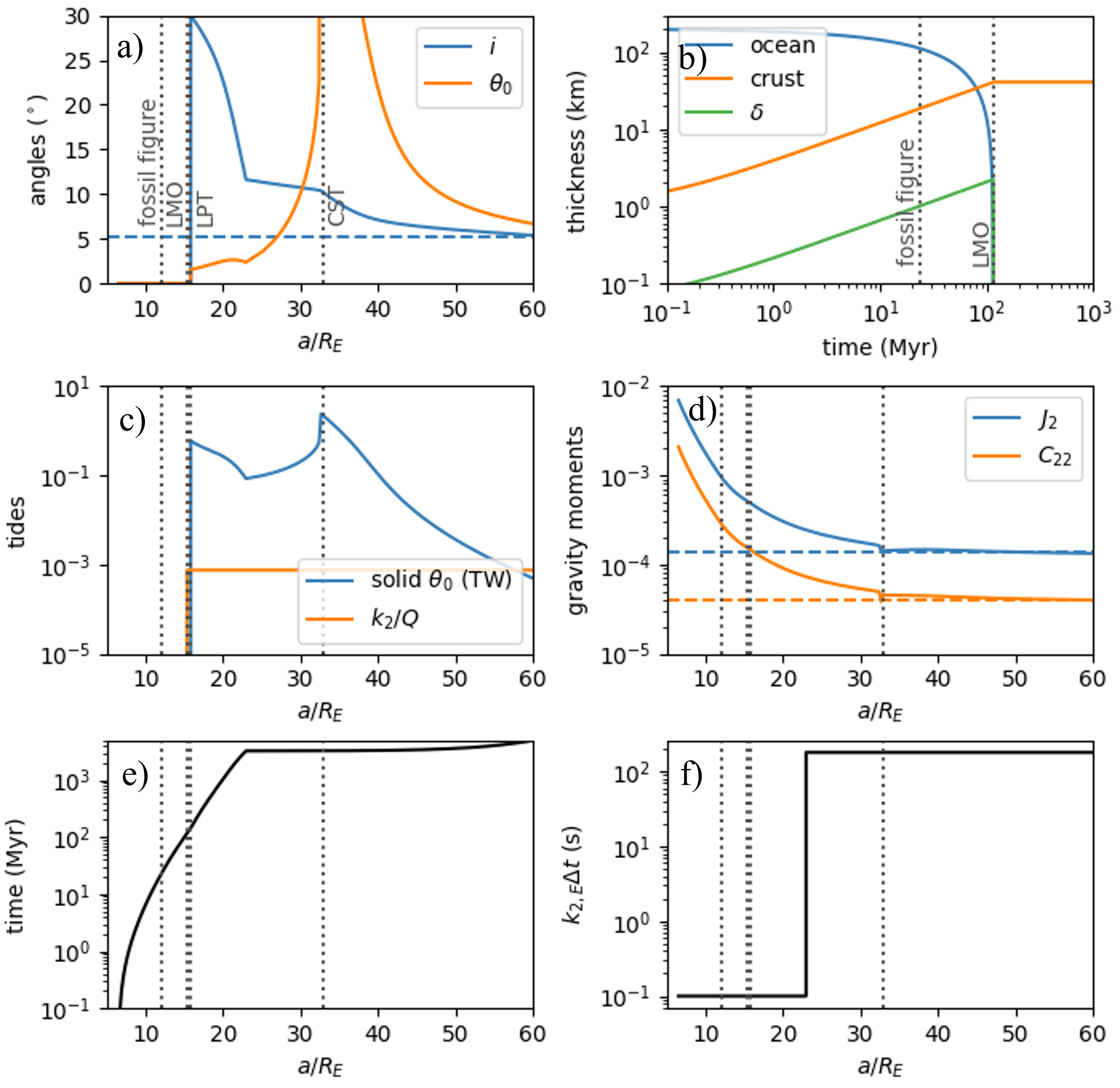}}%
        \caption{Different aspects of an example thermal-orbital evolution. The LPT excites the inclination to $30^\circ$ and the degree-2 gravity observations \citep{matsuyama2021lunar} are reproduced, including a) the inclination and magnitude of the obliquity (negative before the CST and positive after) b) the magma ocean solidification c) solid-body obliquity tidal heating and $k_2/Q$ d) the degree-2 gravity coefficients e) the semi-major axis and f) $k_{2,E}\Delta t$. The magma ocean crystallizes within 115~Myr ($\kappa = 5 \times 10^{-7}$~m$^2$/s) at $a=15~R_E$ for $(k_{2,E}\Delta t)_0$ = 0.1~s. A fossil figure freezes in at $a^* = 12 R_E$, which pushes the CST out to $\sim 33 R_E$ at 3.3~Gyr. The bottom drag coefficient is the value for oceans on Earth, $c_D = 0.002$. Vertical dotted lines mark the fossil figure freezing, magma ocean solidification (LMO), LPT, and CST. The horizontal dashed lines are the observed present-day values for comparison with their solid-color model counterparts.}%
    \label{fig:Moon-run}%
\end{figure}

An example evolution of the LPT exciting the lunar inclination is shown in Fig.~\ref{fig:Moon-run} for $(k_{2,E}\Delta t)_0=0.1$~s. The LPT occurs at $a\sim 16 R_E$ (140~Myr into the simulation) at which point the inclination is instantaneously increased to $30^\circ$ to approximate the results of \citet{cuk2016tidal,cuk2021tidal}. Because the magma ocean solidifies earlier at 115~Myr, only solid-body obliquity tidal heating damps the inclination. We use a constant $k_2/Q=7.5\times 10^{-4}$ that is meant to be an average over the past since the present value is lower and the past value would have been higher. A fossil figure freezes in at $a^* = 12 R_E$ (23~Myr into the simulation), which pushes the CST out to $33 R_E$ (3.3~Gyr into the simulation) compared to $\sim 28 R_E$ had there been no fossil component. The degree-2 gravity coefficients decay as the Moon migrates away from the Earth because the rotation slows and the tidal pull from the Earth weakens. At 3.25~Gyr (1.25~Ga) $k_{2,E}\Delta t$ steps from the initial to the present value to account for the resonant tidal dissipation in the Earth's oceans at the present-day while also getting the Moon to 60~$R_E$ by the end of the simulation. This causes a qualitative change in the semi-major axis evolution at $\sim 23 R_E$ which has a ripple effect on the inclination, obliquity, and tidal heating evolution as well. The knee that this produces in the inclination evolution is by no means critical to the survival of the inclination; it simply reflects the fact that it takes the Moon 3~Gyr in this simulation to go from the LPT at $16 R_E$ to $23 R_E$ and $\sim 1$~Gyr to go from $23 R_E$ to $60 R_E$. Solid-body tidal heating decreases with distance and increases with obliquity at the CST, but the net effect is that more inclination will damp in the first 3~Gyr than in the last 1~Gyr.

\section{Discussion}
\label{Sec:discussion}
The goal of this work is to build on past thermal-orbital models to constrain the evolution of the lunar inclination. We highlight how solid-body tides affect inclination damping, what problems remain for the eccentricity evolution, and how the lunar figure model from \citet{matsuyama2021lunar} allows us to pinpoint when the fossil figure was established and when the CST occurred.

\subsection{Solid-body tides}
A main difference between this work and \citet{chen2016tidal} is that we include solid-body tidal heating for both inclination and eccentricity damping. In Sec.~\ref{Sec:early inc} we show that without ocean tidal heating, solid-body tides can damp away an early inclination of 12$^\circ$, and in Sec.~\ref{Sec:late inc} we show that solid-body tidal heating during the CST could damp away excitations from plantesimals. This is even despite the fact that obliquity tides are stronger in the oceans of synchronous satellites than in the solid-body \citep[e.g.,][]{chen2014tidal}. The results are in line with \citet{cuk2016tidal} who found that the initial inclination of the Moon had to be $>50^\circ$ to decay to its present value assuming the Moon's current solid-body dissipative properties. The lunar evolution adage that the Moon had to begin with twice the inclination it has now \citep[e.g.,][]{goldreich1966history} neglected the importance of solid-body obliquity tides during the Moon's CST.

The opposite is true for eccentricity tides, which are typically stronger in the solid-body of synchronous satellites than in global oceans \citep[e.g.,][]{chen2014tidal}, so solid-body tidal heating is the main driver of eccentricity damping. What we find is that given $k_2/Q=6\times 10^{-4}$ and the fact that solid-body tidal heating is stronger at smaller semi-major axis, slow lunar recession causes any initial eccentricity or excitation at the LPT to damp away completely. With fast recession and arbitrary eccentricity kicks from planetesimals, an eccentricity can survive damping and even increase due to tides on the Earth. Future work on the Moon's orbital evolution will have to focus on simultaneously recovering the Moon's inclination and eccentricity since they may require different explanations. In particular, while we favour the LPT for the Moon's inclination, collisionless planetesimal encounters may well provide an explanation for its eccentricity.

\subsection{Fossil figure}
\begin{figure}
    \centering
{\includegraphics[width=0.55\columnwidth]{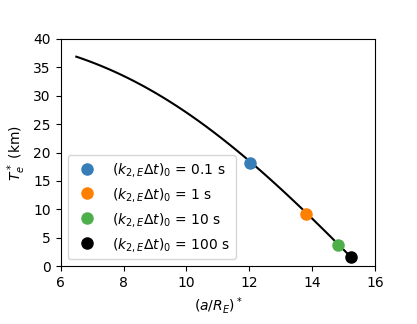}}%
        \caption{The theoretical relationship between the elastic lithosphere thickness, $T_e^*$, and the semi-major axis, $a^*$, when the fossil figure is established to reproduce the present-day lunar fossil figure \citep{matsuyama2021lunar}. Successful simulations with different $(k_{2,E}\Delta t)_0$ are plotted as circles; the blue circle is the model run shown in Fig.~\ref{fig:Moon-run}. There is a balance between the strength of the elastic lithosphere (depends on $k_2^{\infty*}$ or $T_e^*$ by proxy) and the tidal-rotational potential (depends on $a^*$ and assumes negligible obliquity and eccentricity). If the Moon recedes slowly, a thick lithosphere coincides with a short distance, and if the Moon recedes quickly, a thin lithosphere coincides with a large distance. There is an upper bound of $a^*=16 R_E$, which corresponds to an elastic lithosphere thickness of 1~km ($k_2^{\infty*}=1.359$). The results assume that the magma ocean did not form a permanent flotation crust until 6.5 $R_E$.}%
    \label{fig:d-freeze}%
\end{figure}

The size of the fossil figure observed at the present-day depends on two factors, the strength of the elastic lithosphere and the strength of the tidal-rotational potential when it was established (see Eqs.~\ref{Eq:J2_fossil} and ~\ref{Eq:C22_fossil}). Striking a balance between the two means considering the relative timing of the solidification of the magma ocean with respect to the semi-major axis migration. Fig.~\ref{fig:d-freeze} shows the theoretical relationship between the elastic lithosphere thickness, $T_e^*$, and the semi-major axis, $a^*$, when the fossil figure is established in order to produce the fossil figure size in \citet{matsuyama2021lunar}. The more resistant the lithosphere is to deformation, the stronger the tidal-rotational potential has to be (large \textcolor{orange}{$T_e^*$}, small $a^*$). The more easily the lithosphere deforms, the weaker the tidal-rotational potential can be (small $T_e^*$, larger $a^*$). To achieve a thick elastic lithosphere at a small distance, the Moon must recede slowly due to a small $(k_{2,E}\Delta t)_0$. To achieve a thin elastic lithosphere at a larger distance, the Moon must receded quickly due to a large $(k_{2,E}\Delta t)_0$. 

Fig.~\ref{fig:d-freeze} shows that there is a theoretical upper bound of $16 R_E$ for freezing in the fossil figure, which coincides with an elastic lithosphere thickness of 1~km. For $(k_{2,E}\Delta t)_0 \le 0.1$~s consistent with the LPT exciting the lunar inclination (blue dot), the fossil figure freezes in at $\le 12 R_E$. These predictions are noteworthy because it would mean that the fossil figure predates both the LPT at $\sim 16 R_E$ and the CST at $\sim 33 R_E$. Thus, an important conclusion from our work is that neither event could have heated the Moon strongly enough to reset the lunar figure. More sophisticated coupled thermal-orbital models could test this conclusion. Furthermore, for the $(k_{2,E}\Delta t)_0$ sampled in Fig.~\ref{fig:CST-LPT}, the fossil figure always freezes in prior to 100~Myr after Moon formation. With more consensus on the ages of the Moon and the South Pole-Aitken Basin (SPA), more could be said about the relative timing between the fossil figure and SPA.

There are a few effects that we have ignored for simplicity that could lead to over- and under-estimating the fossil figure contributions to the degree-2 gravity field. In this work, we approximate the elastic lithosphere thickness as being the total crustal thickness when in reality it is smaller. The consequence is that our simulations potentially freeze in the fossil figure before the elastic lithosphere has reached its optimal strength, and so the curve in Fig.~\ref{fig:d-freeze} should be pushed to the right to larger semi-major axes to correct for this. Another factor that could lead to overestimating the fossil figure is that we neglect strengthening of the elastic lithosphere as the Moon cools. Strengthening the elastic lithosphere lowers the long-term Love numbers and causes the fossil figure contribution to grow with time. For reference, our simulations have a maximum of $T_e^=45$~km, which corresponds to $k_2^{\infty}=0.993$. A way that our model underestimates the fossil figure contribution is that it freezes in the fossil component instantaneously and does not account for later viscous relaxation \citep[c.f.,][]{qin2018formation}. For a long-term formation of the fossil figure as in \citet{qin2018formation}, as long as the bulk of the fossil figure is established prior to the CST, the effects on our results would be limited. To freeze in a larger fossil figure than that observed today in order for the stresses to viscously relax over time would push the black curve in Fig.~\ref{fig:d-freeze} to the lower left, meaning that the fossil figure would have to be established at smaller semi-major axis and elastic lithosphere thickness.

In all of the simulations that reproduce the degree-2 gravity observations, the CST happens in the range of $\sim 32-38 R_E$, which is consistent with the 34$R_E$ found in \citet{matsuyama2021lunar}. We show in Sec.~\ref{Sec:late inc} that without a fossil figure, the CST would happen at $\sim 28 R_E$. The observation that a fossil bulge affects where the CST occurs was also observed by \citet{siegler2011effects} in addition to \citet{matsuyama2021lunar}. In \citet{siegler2011effects}, the lunar figure has a combination of hydrostatic and fossil components, although in what proportion is not stated. Their CST happens at $\sim 30 R_E$. This is compared to $\sim 28 R_E$ for the absent fossil figure case in this work and to $33 R_E$ with a fossil figure. \citet{ward1975past} also calculated a CST at $\sim 34 R_E$ but that was assuming the present-day lunar figure and orbit. In reality, the figure and orbit evolved significantly enough in the past to affect the location of the CST, so it is a coincidence that \citet{ward1975past} is consistent with our results.

\section{Conclusion and Future Work}
\label{Sec:conclusion}
We conclude that the most likely source of the lunar inclination is the LPT at $a = 16 R_E$ \citep{cuk2016tidal, cuk2021tidal}, and for this point to be reached after the solidification of the magma ocean in $>100$~Myr, the Moon recedes slowly from an early Earth with $(k_{2,E}\Delta t)_0 \le 0.1$~s. The other possibility is that collisionless encounters with planetesimals in the first 140~Myr after Moon formation \citep{pahlevan2015collisionless} excite the lunar inclination, which requires quickly passing the CST and solidfying the magma ocean before this time with ($k_{2,E}\Delta t)_0 \ge 40$~s. A large $(k_{2,E}\Delta t)_0$ is inconsistent with a weakly dissipative Earth soon after Moon formation \citep{zahnle2015tethered}. For slow initial recession from the Earth, the lunar fossil figure was established at $\lesssim 12 R_E$. The eccentricity and obliquity during this time are small compared to other fossil figure solutions that require a large eccentricity or non-synchronous rotation. Our fossil figure range is consistent with that found in \citet{matsuyama2021lunar}. 

Future work on the question of how the Moon got its inclination and where the fossil figure was established will entail including the conditions necessary for the LPT to excite the lunar inclination as in \citet{cuk2016tidal,cuk2021tidal}, the effect of topographic variations on ocean tidal heating, and the early tidal state of the Earth. Since the location of the LPT is dependent on the angular momentum of the early Earth-Moon system, future work should explore the effects of a high angular momentum starting point on the predicted tidal-orbital lunar evolution. Our ocean tides model assumes a uniform ocean thickness with no crustal variations. To account for some non-uniform ocean solidification, we turn off tidal heating in the ocean when the thickness reaches 10~km. \citet{rovira2020tides} showed that the effects of ocean thickness variations on tidal heating are much more complicated and have the potential to limit tidal heating considerably. In future work, we will see how potential variations in the ocean thickness can affect the obliquity tide flow. 

Modelling the thermal-orbital evolution of the Moon would be improved with a better understanding of the tidal state of the early Earth. Our simple parameterization of $k_{2,E}\Delta t$ recovers the correct present-day lunar semi-major axis and Earth $Q$ but is {\em ad hoc} and could in principle be improved if we had sufficient understanding of ocean dynamics throughout Earth history. The early tidal state of the Earth is key to ruling out different lunar evolution scenarios in our coupled thermal-orbital model.

\begin{table}
\centering
\caption{Physical parameters used in the thermal-orbital model}\label{Tab:params}
\begin{tabular*}{\tblwidth}{@{}LL@{}LL@{}}
\toprule
Symbol & Parameter & Value \\ 
\midrule
 $G$ & Gravitational constant & 6.674$\times 10^{-11}$ m$^3$ s$^{-2}$ kg$^{-1}$ \\
 $M_S$ & Mass of the Sun & 2$\times 10^{30}$ kg \\
 $M_E$ & Mass of the Earth & 5.9723$\times 10^{24}$ kg \\
 $R_E$ & Radius of the Earth & 6378 km \\
 $a_E$ & Semi-major axis of the Earth & 1.496$\times10^{11}$ m \\
 $k_{2,E}$ & $k_2$ Love number of the Earth & 0.97 \\
 $\alpha$ & Normalized moment of inertia of the Earth & 0.33 \\
 ${m}$ & Mass of the Moon & 7.25$\times 10^{22}$ kg \\  
 $R$ & Radius of the Moon & 1737 km \\
 $\rho_o$ & Density of the magma ocean & 3000 kg m$^{-3}$ \\
 $C_p$ & Specific heat capacity of magma ocean & 1256 J kg$^{-1}$ K$^{-1}$ \\
 $T_m$ & Temperature of magma ocean & 1200 K \\
 $T_0$ & Temperature of the surface & 280 K \\
 $T_a$ & Temperature of layer interface in crust & 1150 K \\
 $L$ & Latent heat of fusion & 5$\times 10^5$ J kg$^{-1}$ \\
 $c_D$ & Bottom drag coefficient & 0.002 \\
 $\eta_o$ & Viscosity of magma ocean & 1000 Pa s \\
 $\mu$ & Rigidity of the crust & 30 GPa \\
 $\eta$ & Viscosity of the crust & $10^{19}$ Pa s \\
 $f$ & Spatial variable & 1 \\
\bottomrule
\end{tabular*}
\label{tab:Moon params}
\end{table}

\section*{Acknowledgements}
We thank two anonymous reviewers for their thorough comments and suggestions which have significantly improved this manuscript. I. M. was supported by the National Aeronautics and Space Administration (NASA) under grant No. 80NSSC20K0334 issued through the NASA Emerging Worlds program. This material is based upon work supported by the National Science Foundation Graduate Research Fellowship under Grant No. DGE-1842400.

\section*{Declaration of Competing Interest}
The authors declare that they have no known competing financial interests or personal relationships that could have appeared to influence the work reported in this paper.

\bibliographystyle{cas-model2-names}

\bibliography{moon.bib}

\end{document}